\documentclass[12pt]{iopart}
\usepackage[pdftex]{graphicx}
\newcommand{\beq}{\begin{equation*}}
\newcommand{\eeq}{\end{equation*}}
\newcommand{\beqa}{\begin{eqnarray*}}
\newcommand{\eeqa}{\end{eqnarray*}}
\begin{document}

\title[Dark Matter Frontier]{Cosmology and the Dark Matter Frontier}

\author{Lars Bergstr\"om}

\address{The Oskar Klein Centre, Department of Physics\\
Stockholm University, AlbaNova University Center\\
 SE-106 91 Stockholm, Sweden}
\ead{lbe@fysik.su.se}
\begin{abstract}
A brief overview is given about some issues in current astroparticle physics, focusing on the dark matter (DM) problem, where the connection to LHC physics is particularly strong. New data from the Planck satellite has made the evidence in favour of the existence of DM even stronger. The favourite, though not the only, candidates for cosmological DM, weakly interacting massive particles (WIMPs), are being probed by a variety of experiments - direct detection through scattering in terrestrial detectors, indirect detection by observing products of annihilation of DM in the Galaxy, and finally searches at accelerators such as the LHC. The field is in the interesting situation that all of these search methods are reaching sensitivities where signals of DM may plausibly soon be found, and a vast array of models will be probed in the next few years. Of course, expectations for a positive signature are high, which calls for caution regarding ``false alarms''. Some of the presently puzzling and partly conflicting pieces of evidence for DM detection are discussed as well as expectations for the future.   

\end{abstract}

\maketitle

\section{Introduction}
We have seen in many contributions at this Symposium that there is an intense struggle to find even small dents in the overwhelmingly successful Standard Model of particle physics. With a Higgs boson just discovered \cite{atlas,cms}, the thought may appear that this was the last discovery for a very long time, as it makes the Standard Model qualitatively complete (of course it does not explain some small numbers, like neutrino masses or the value of the cosmological constant, but these values can be adjusted by hand in the theory). Fortunately, for those of us who think that much remains  to be discovered, the Standard Model of cosmology, with the acronym the $\Lambda$CDM model, provides strong arguments in favour. This is particularly the case for the existence of dark matter (DM, or often CDM, for ``cold'' dark matter, i.e., slowly moving massive particles) -- something that definitely leads to physics beyond the Standard Model. ($\Lambda$ is Einstein's cosmological constant.) 

The basic features of the $\Lambda$CDM model are very simple (see, e.g., \cite{lbe_goobar}). We assume a homogeneous and isotropic average background geometry of the universe (as indicated, for instance, by the smallness of the fluctuations in the cosmic microwave background, CMB). Writing for the general relativistic metric
\beqa ds^2=dt^2-a^2(t)\left({dr^2\over 
 1-kr^2}+r^2d\theta^2+r^2\sin^2\theta 
 d\phi^2\right),\eeqa
the only non-trivial quantities appearing are the scaled curvature $k$ and the scale factor $a(t)$, from which which the expansion rate $H(t)$ is defined by 
$$H(t)^2=\biggl(\frac{\dot a(t)}{a(t)}\biggr)^2.$$ 
The constant $k$ is related to the overall spatial curvature of the universe, $k=0$ for a flat universe, $k=1$ for a closed and $k=-1$ for an open universe. The overall geometry of the universe can also be specified by the value of $\Omega_{tot}$, i.e. the total energy density $\rho_{tot}(t_0)$ today, at time $t_0$, divided by the critical density of a flat universe, $$\Omega_{tot}=\frac{\sum_i\rho_{i}(t_0)}{\rho_{crit}(t_0)}.$$
Here $i$ sums over baryons, DM and dark energy (plus a radiation component that is very small today, but which was much bigger, or in fact dominant, in the early universe). 
One of Einstein's equations, the Friedmann equation, gives the connection between the expansion rate and the total energy density, 
$$
H(t)^2=\biggl(\frac{\dot a(t)}{a(t)}\biggr)^2=\frac{8\pi G}{3}\sum_i \rho_i(t)-\frac{k}{a(t)^2}.$$ 
Putting $k=0$, i.e. for a geometrically flat universe, we see that the critical value of the total density today, at time $t_0$, is
$$\rho_{crit}=\frac{3H_0^2}{8\pi G},$$
with $G$ Newton´s constant, and $H_0=H(t_0)$ the Hubble constant (i.e., the present value of the Hubble parameter). 
Models of cosmic inflation predict $\Omega_{tot}$ to be very close to 1, and this in fact vindicated by CMB data, where, writing $\Omega_K=1-\Omega_{tot}$, recent Planck satellite data \cite{planck_inflation} supplemented by the pioneering WMAP satellite large-scale polarization likelihood \cite{wmap_pol} and baryon acoustic oscillations (BAO) \cite{bao} implies
\beq
\Omega_{K} = -0.0004\pm 0.00036.
\eeq 
This corresponds to such a small deviation from a flat universe, that usually CMB and other cosmological data are analyzed under the assumption that $k=0$.

The amount of baryonic matter in the universe $\Omega_B$, is given from two sources of information, the analysis of CMB, and from primordial nucleosynthesis which give consistent results. The most accurate, from Planck 2013 (including also WMAP polarization data) is
\beq
\Omega_Bh^2=0.02205\pm 0.00028.
\eeq

The Planck Collaboration also gives an accurate estimate of the amount of cold DM, CDM
\beq
\Omega_{CDM}h^2=0.1199\pm 0.0027,
\eeq 
where $h$ is the scaled Hubble constant, $h=H_0$/(100 kms$^{-1}$Mpc$^{-1})$. The best fit given by Planck is $h=0.673 \pm 0.012$, somewhat smaller than the previous best value (for example, WMAP 7-year value \cite{wmap}, $h=0.704\pm 0.025$). One finds, dividing out $h^2$, the central value $\Omega_{CDM}=26.4 \%$, and similarly $\Omega_B=4.9 \%$. With $\Omega_{tot}=1$, one then finds for the contribution to the energy density from the cosmological constant $\Lambda$, $\Omega_\Lambda=68.7 \%$, with an uncertainty of a few percentage units. 

This  shows the present challenge involved in understanding cosmology: What are the agents that make up 95 $\%$ of the energy density in the universe? Still today, we do not have many clues about the value of $\Lambda$ which in the natural units of gravity (i.e. defined by the only mass scale, the Planck constant) naively seems to be some 120 orders of magnitude too small, but still non-zero. It seems that ``anthropic'' reasoning \cite{weinberg} is presently as at least as good as any traditional scientific method to explain the measured value. 
 
A novel feature of the Planck data is the measurement of gravitational lensing of the microwave background radiation on its way to us \cite{planck_lensing}. The main part of the gravitational lensing signal comes from redshifts around 2 - 3, when galactic structure has formed. Distortions of the angular spectrum due to gravitational lensing of these large scale structures has been detected with high significance and the distribution fits well with the predictions of the $\Lambda$CDM model with parameters given above, giving yet another strong indication for the need for DM.

Other important pieces of information from the recent Planck data, obtained from a fit to the angular power spectrum supplemented by the cosmological data is a limit on the sum of neutrino masses which ranges from around 0.3 eV to 1 eV, depending on the details of the analysis. Writing in terms of the effective number of neutrinos (which may also include other, very light particles), the Planck result is $N_{eff}=3.30^{+0.54}_{-0.51}$, which is consistent with only the three known light neutrinos species contributing, although there is still some room for extra effective degrees of freedom.

To conclude, the $\Lambda$CDM model, sometimes named the ``concordance model'' has so far  survived a multitude of different consistency checks, and its parameters have been accurately determined. One of the big unknown questions today, which gives a strong connection between cosmology and particle physics, is that of the composition of DM, which constitutes a fourth of the energy density - five times more than visible matter, baryons.
\section{Dark Matter}
Let us try to investigate dark matter from the point of view of Standard Model particles. To describe matter that has survived in the universe until this day, we obviously need particles with a very long lifetime, much longer than the age of the universe, which according to the Planck measurements is 13.8 billion years. We also need a rather heavy particle (at least heavier than a few keV) to fulfill the requirement of moving slowly at the time of decoupling from the thermal heat bath in the hot early universe, i.e., to be cold DM. The particle also has to be electrically neutral, (or have a very small, non-standard charge) since otherwise it would not be dark - it would radiate electromagnetically (by the way, ``invisible matter'' would really be a better word, but the term introduced by F. Zwicky in 1933 \cite{zwicky}, ``dark matter'' is the one that has stuck.) 

It is easy to see that the Standard Model (SM) of particle physics contains no suitable candidate. The $Z^0$ boson and the recently discovered Higgs boson $H^0$ are massive and neutral, but have a lifetime of a fraction of a second only, so are not useful for forming any structure. The only known stable baryon is the proton which however has non-zero charge and is thus not invisible. 
The electrically neutral neutron decays within some 10 minutes if free, and if bound in electrically charged nuclei would count as visible (baryonic) matter, not CDM.

Finally, as we have seen, the three known neutrinos, although postulated to be massless in the SM, in any case have a mass below 1 eV, which would make them hot and not cold DM, i.e., they moved relativistically when structure formed -- something that is strongly disfavoured by observations.

We thus have to look beyond the Standard Model to find suitable DM candidates, and at least we know what we are looking for: massive, electrically neutral particles with long lifetime. These candidates are called WIMPs (Weakly Interacting Massive Particles). If we put more faith in standard cosmology, we may also try to find models with massive stable particles that give the measured $\Omega_{CDM}h^2$. 

The physics is again simple: To ensure stability, assume a conserved quantum number, the simplest being a parity-like $Z_2$ symmetry. During the first fractions of a second after the big bang, WIMPs were continuously created, and annihilated in pairs, to particles making up the primordial heat bath, until the expansion of the universe made the number density to small for WIMP pairs to encounter each other. As they are assumed to be stable, this means that the number density in comoving coordinates still today is very close to constant. The relic density computed in this way from early universe thermodynamics gives a very simple result (see the pioneering paper \cite{lee} and  any of the reviews \cite{jkg,lbereview,bertonesilk,bertone_book,feng} for more details):

\beq
\Omega_{WIMP} h^2\simeq 0.12\cdot
\left(\frac{2.6\cdot 10^{-26}\ {\rm cm}^3{\rm s}^{-1}}
{\langle\sigma_A|{\mathbf v}|\rangle}\right),
\eeq
where $\langle ...\rangle$ denotes taking the average of the relative velocities and angles (this average
is often just abbreviated as  $\sigma v$). In many plots showing DM predictions and limits, one usually displays this ``WIMP relic density'' value, $\sigma v\sim 3\cdot 10^{-26}\ {\rm cm}^3{\rm s}^{-1}$, which interestingly does not depend explicitly on the WIMP mass. The fact that an annihilation cross section of a picobarn, i.e., a typical weak interaction cross section, gives the observed $\Omega_{WIMP}h^2\sim \Omega_{CDM}h^2 = 0.12$ may be of significance (sometimes it is called the ``WIMP miracle''), and WIMPs have become the leading DM candidates.

As many WIMP candidates are Majorana particles, there is no particle-antiparticle asymmetry. Of course, some model builders are more ambitious and want to find a connection between the baryon asymmetry in the universe and dark matter. This leads to models of asymmetric dark matter  (see, e.g., the review \cite{zurek})
with  somewhat different phenomenology. 

\section{Supersymmetric and other DM Candidates}
As mentioned, a simple way to ensure stability of a DM candidate is to
assume a discrete $Z_2$ symmetry, such that all Standard Model particles have positive
$Z_2$ parity and particles in the ``dark sector'' have negative parity. Then, the lightest particle with 
negative $Z_2$ parity, if it is electrically neutral, is an attractive DM candidate, as its stability
is guaranteed by the $Z_2$ parity. This is in fact
how the most studied DM candidate -- the lightest supersymmetric particle -- appears in theories of 
supersymmetry. In  supersymmetric models that are viable from the DM point of view, there is indeed
a multiplicatively conserved $Z_2$ quantum number, $R$-parity, which is $+1$ for 
all ordinary particles of the Standard Model and $-1$ for supersymmetric 
partners. In terms of baryon number $B$, lepton number $L$ and and spin $s$ it
can be written 
$R=(-1)^{3(B-L)+2s}$.

In supersymmetric models, the conservation of $R$-parity also forbids otherwise 
problematic large baryon number violating couplings to occur. As DM has to be electrically neutral, a good candidate is the lightest neutralino (which is in general a quantum mechanical mixture of the supersymmetric partners of the photon, the $Z^0$ and the two neutral scalar Higgs bosons needed in supersymmetric models). From the number of fermion degrees of freedom required by supersymmetry, one can further show that neutralinos have to be Majorana fermions, i.e., they are their own antiparticles.

Of course, we do not 
know if supersymmetry exists. As we have heard at this Symposium, so far there is no hint of its existence
 at LHC -- although we are still at an early stage of LHC operation, not yet at design energy.
At least some of the proponents of supersymmetric DM (see the view \cite{ellis} of one of the 
earliest inventors, \cite{ellis2}) still hope that it will
be discovered in the 13-14 TeV runs that start in the next couple of years. It remains to be seen, though,
if the simplest supersymmetric scenarios, now put under some pressure as gluinos and squarks have not yet been detected, will survive fortcoming LHC data. Also,
a difficulty in these ``constrained'' supersymmetric models may be to explain the large radiative corrections
needed to give the rather high
Higgs mass 126 GeV, to be compared with the lowest order prediction $m_H < m_Z$ in minimally supersymmetric models.
 However, the example of $R$-parity conservation as mechanism of creating good DM candidates 
by having a $Z_2$-like parity symmetry is very general and is one of the key 
ingredients of many, if not most, WIMP DM
models. 

Irrespective of its present lack of experimental confirmation, the lightest supersymmetric particle is 
still an excellent WIMP template, with detection properties that are 
completely computable once the (many)
parameters of the model have been specified (e.g., using the {\sc DarkSUSY} package \cite{ds}, which is presently being updated to serve as a more model-independent tool).

Among alternate models can be mentioned those with  universal extra dimensions, Kaluza-Klein (KK) models (see \cite{profumo} for a review),
where there a ``KK-parity'' appears, which has the value of $+1$ for Standard Model particles
and $-1$ for the first excited KK states in these models, which again means that the lightest
parity $-1$ particle is stable. Usually, this is an excited, electrically neutral $U(1)$ boson, thus having spin-1, in contrast to the spin-1/2 neutralino of supersymmetric DM. An analysis of present LHC results gives the limit $> 600 -700$ GeV, for the mass scale of these models \cite{ued}. 

One may even construct a minimalistic DM model by just extending the Standard Model (SM) with one extra
singlet or doublet Higgs. In the case of a singlet $S$, one can impose the discrete
symmetry under $S\to -S$ for the potential which couples $S$ to the SM Higgs doublet (e.g., \cite{zee}).  
According to \cite{cline}, this simplest DM model is still viable, although the currently allowed parameter 
space is on the verge of being significantly reduced
with the next generation of experiments.

All these models are examples from the WIMP category. Of course, there are candidates which 
have other motivations from particle physics. The prime example is the axion, introduced to solve
the strong CP problem in QCD \cite{axions}. New experiments (based on the axion to photon coupling \cite{sikivie}) 
are starting to penetrate the remaining 
allowed window \cite{ksvz} which gives the required relic DM density \cite{admx}. Although these axions are very light, much less massive than 1 eV,
they were produced by other mechanisms than the thermal one, and they in fact
behave very much like CDM. 

An idea, which currently is much investigated, is that extra, light U(1) fields, ``dark photons''  could make up the dark matter. Of course there is little direct need for such additions to the SM, but the phenomenology is interesting (see \cite{dark_photons} for a review).

The methods to search for axions and other very light particles are quite different from the ones used to search for WIMPs, to which we now turn.

\section{WIMP DM Detection Methods}
There are in broad terms three different methods employed when trying to detect DM particles such as WIMPs, and estimate their mass and interaction strength (actually, there could also be particular astrophysical effects like an influence on stellar evolution, such as ``dark stars'' \cite{darkstars}, but the situation is not clear about their observability.)

The first method, {\em Direct detection} of DM, relies on the fact that DM particles should omnipresent in the universe, in particular it should be present in the  halo of DM in which the Milky Way is embedded. DM particles should typically move with normal galactic velocities (as measured, e.g., from the rotation curve) 
of $\sim 200$ km/s, i.e., with $v/c\sim 10^{-3}$. From modeling of the Milky Way 
one finds that locally the mass density of DM should be of the order of $0.4$ GeV/cm$^{3}$,  roughly a factor of 
two uncertainty (see \cite{catena}, where, assuming a spherically symmetric halo,  the uncertainty is given as low as 7 \%). Combining these numbers one can estimate that, if the cross 
section is that of a supersymmetric WIMP, scattering on nuclei should take place with a cross section
between a few times $10^{-4}$ pb to $10^{9}$ pb \cite{ds}, but with large uncertainties (see \cite{goodman} for an early paper on the general idea behind direct detection). In deep underground laboratories, 
there are a number of experiments taking data or being deployed, with XENON-100 presently giving the 
best bounds \cite{xenon100} (see also CDMS-II \cite{cdms}).

Direct detection experiments have been rapidly evolving 
 in the mass range from around 30 GeV to a TeV where they nicely 
complement accelerator searches. 
In these detectors, 
rare events giving a combination of scintillation, 
ionization, phonon or total energy deposition signals are searched for, in noble gas or solid state detectors, shielded 
from cosmic rays in underground sites.  The situation regarding present data is somewhat confusing. For a long time, the DAMA/LIBRA experiment in the Gran Sasso Laboratory has been showing 
evidence (presently reaching 9$\sigma$) of an annual modulation signal \cite{dama}. This is expected in dark matter models (see  \cite{freese}) due to a different DM ``wind'' direction as the earth moves around the sun, combined with the strong relative velocity dependence of the scattering rate of dark matter. The origin of the observation of annual modulation is unexplained, but has  not been confirmed by other experiments \cite{cdms,xenon100,kims12}. The same is true for the 
 CoGeNT excess events and annual modulation \cite{cogent}, which also is in tension with \cite{cdms,xenon100}.  The explanation of these 
experiments in terms of DM scattering seems to probably 
necessitate leaving the standard WIMP scenario, or the use of non-standard dark matter 
halo models and/or fine-tuning the DM coupling to Xenon to be unexpectedly small. The experimental situation is improving rapidly, but recently another indication of DM scattering was presented in silicon data from CDMS-II \cite{cdms-ii}, leaving the situtation even more confusing. Maybe it will take the next generation of experiments, such as XENON-1T (presently being installed) or the planned very large DARWIN experiment to clarify the situation.

The second method is the detection at accelerators of some new class of particles, one of which may be the DM. The DM
particle will itself be essentailly impossible to conclusively identify directly, as the required lifetime has to be demonstrated to be much longer than the age of the universe. In particle detectors, this is not possible to prove in general as the only ``trace'' is missing energy or transverse momentum. 
However, by obtaining a signal which sets the mass scale for events beyond the Standard Model 
one can then search for consistency with  experiments of other types that probe the galactic population of dark matter particles. 

\begin{figure}[htb!]
\centering
\includegraphics[scale=0.4]{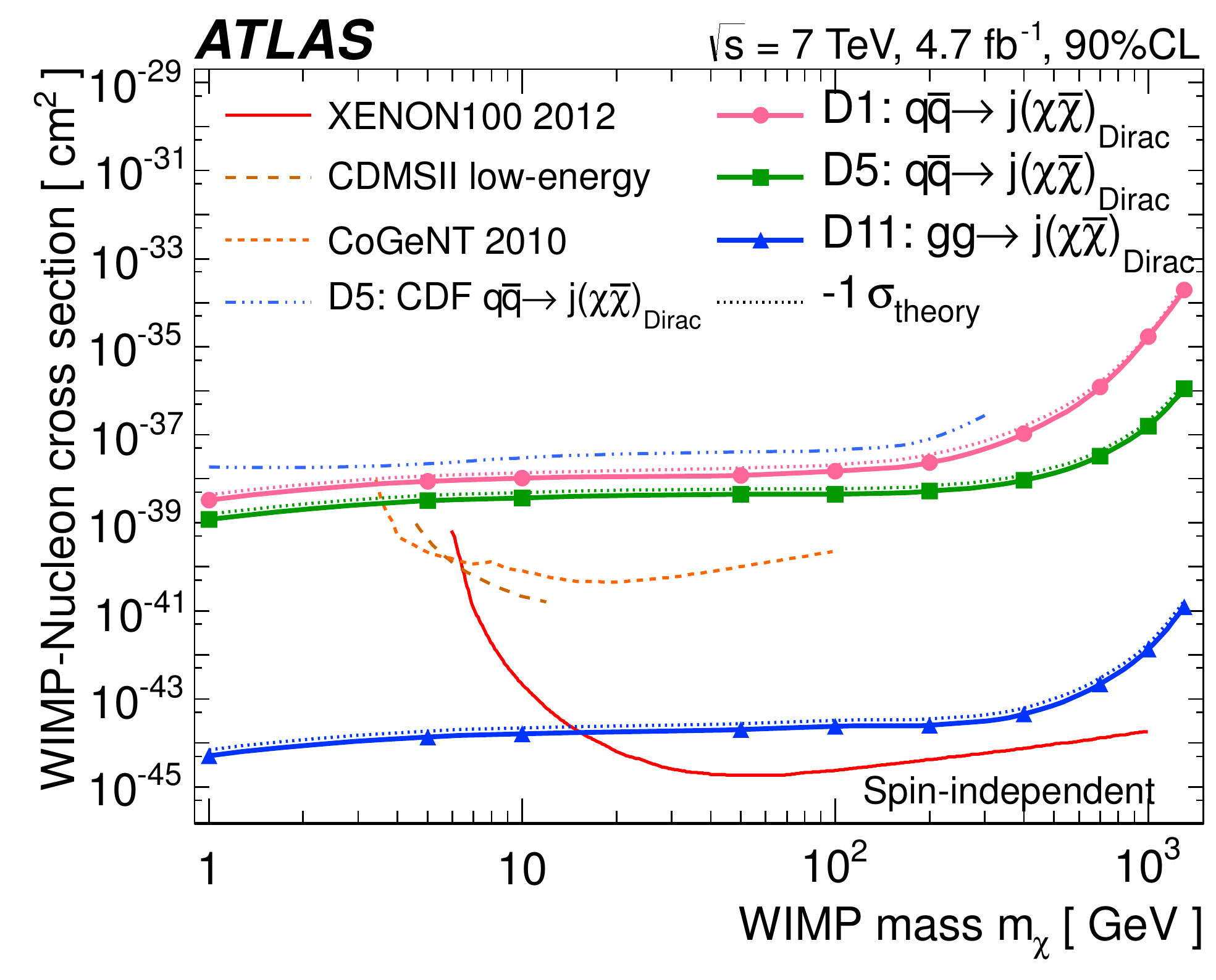}
\caption{\it An analysis using data from the ATLAS collaboration \protect\cite{atlas_jet} showing WIMP-nucleon cross section limits on dark matter candidates with various assumptions for the effective couplings (see \protect\cite{atlas_jet} for details). Also shown are  direct detection limit, which are less sensitive in the region below a few GeV.}
\label{fig:jet}
\end{figure}

An interesting proposal to search for dark matter signals is to search for events with a jet and missing transverse momentum in the detector, A recent result from ATLAS \cite{atlas_jet} is shown if Fig.\ref{fig:jet}. Under some assumptions on the coupling of the dark matter to quarks or gluons with emission of a jet, quite restrictive limits are obtained. Although these first limits are very model dependent, one sees the potential of the method, especially for light WIMPs, where  current direct detectors lose sensitivity. 

In {\em indirect detection}, one  registers products of dark 
matter annihilation from regions in the surrounding universe with a high DM 
density like the galactic centre, dwarf spheroidal galaxies, 
galaxy clusters or even in astrophysical bodies like the sun or the earth (in the last case only escaping neutrinos can give a signal).  
Knowing the halo density distribution, for instance by using numerical simulations (see, e.g., \cite{millennium})
one can estimate the probability for DM particle annihilation into final states containing detectable Standard Model particles. Especially important is the search for  antiparticles like positrons and antiprotons, as these are suppressed in cosmic rays due to the baryon asymmetry of the universe. 

\subsection{Detection by Neutrinos}
 As two particles have to be at the same place to 
annihilate, the annihilation rate  is quadratic in the number density DM particles. Therefore, wherever there is believed to be a density enhancement, an annihilation signal may be searched for. At the smallest scales we find planets or the sun, with estimates indicating that neutrinos from the centre of the sun gives rather strong limits on WIMPs which have a spin-dependent scattering on the protons in the sun. This technique has been pioneered by the IceCube experiment \cite{icecube}. (As the earth contains mainly spin-0 elements, the limits from neutrinos in the direction of  the centre of the earth are not competitive with direct detection limits, for which the spin-independent limits are far superior to the spin-dependent ones.) Depending on the timescale for capture, either the capture rate or the annihilation rate could be the dominant factor. In the annihilations, typically all kinematically allowed Standard Model particles are created. However, only neutrinos would easily
escape (in the case of the sun, this is true at least for DM particles with mass less than a TeV -- if DM is more massive, absorption in the  solar interior would become important). Limits from IceCube \cite{icecube}, complemented by the DeepCore inset,  are competitive for 
spin-dependent scattering (but are for some classes of models being superseded
by those from the LHC \cite{atlas_jet}).

\subsection{Detection through Antimatter}
One of the first suggestions for indirectly detecting dark matter was to use antiprotons \cite{SS}, as the dominant background, secondary antiparticles created in cosmic-ray interactions, are expected to have an energy spectrum which falls rapidly with energy. However, the recently measured antiproton rate by the PAMELA experiment \cite{pamela_pbar} shows no signal of a component above the calculated secondary flux. This important fact can be used to limit the parameters of a number of DM models.

More interesting is the situation concerning positrons. This possible signature also has a long history (see, e.g., \cite{positron_early}), and was put in focus a few years ago when the PAMELA experiment \cite{pamela_pos} showed that the fraction of positrons was rising with energy, at least up to 100 GeV, contrary to the expectations from the model with secondaries only \cite{moskalenko_strong}. Soon thereafter, also the Fermi satellite reported an excess in the electron plus positron flux at a few hundred GeV \cite{fermi_pos}. This caused a flurry of theoretical activity, with many proposals for DM models to explain the positron excess. However, although one could tweak the theoretical explanations so as to fit the measured spectra, one had to go rather far beyond the standard WIMP scenarios. First of all, as nothing is seen in the antiproton spectrum, viable models have to be ``leptophilic'', i.e., has to couple very weakly to quarks and gauge bosons. Secondly, one has to postulate a very large annihilation rate, meaning that large ``boost factors'' are needed (see, e.g., \cite{zaharijas}). Although none of these facts are fatal for models builders (one may, for instance have an enhancement due to an effect analogous to the one first found in electrodynamics by Sommerfeld \cite{arkani}), these models may appear contrived.

Very recently, the AMS-02 experiment on the International Space Station has presented excellent data on the positron flux ratio up to 350 GeV, which confirms the rising positron fraction. With further data, maybe also a fall-off can be detected, as expected in DM models. However, positrons are very light and thus are easily produced in a number of astrophysical processes, such as pulsars (see, e.g., \cite{grasso}). The lack of a specific difference in signature between DM and pulsars will make the discrimination difficult (the expected directional asymmetry if one nearby pulsar dominates seems to be to small for present experiments to detect). On the other hand, the excellent quality of the AMS-02 data at low energies, which follows well a smooth parametrization, may be used to put stringent limits on conventional WIMP DM models (with boost factors of order unity) that would give a ``bump'' in the spectrum, such as those that annihilate directly into an electron-positron pair \cite{lbe_hooper}.

\subsection{Gamma-ray Detection}
Recently, many experiments of indirect detection have been dealing with $\gamma$-rays from DM annihilation. This is, similarly as for neutrinos, due to several unique properties of $\gamma$-rays.
First of all, they do not scatter appreciably during their travel through the galaxy, but rather point back at the site where the annihilation took place. Also, absorption can generally be neglected, as the cross section for scattering on electrons and nuclei for the GeV to TeV range is very small. This means that one may use properties of the energy distribution resulting from these processes to separate a signal from astrophysical fore- or backgrounds. And of course, as the electromagnetic cross section of $\gamma$-rays is so much higher than the weak interactions for neutrinos, they are relatively easy to detect.

The photon energy is limited from above by the rest mass of an annihilating particle. This comes from energy- and momentum conservation, and the fact that WIMP DM moves today moves non-relativistically in the universe.  A currently interesting case is the annihilation into two photons, give a monochromatic line at energy $E_\gamma=m_X$, with $X$ being the DM particle.
(This is a ``smoking gun'' process first suggested in \cite{lbe_snellman}). 
It is however loop-suppressed by a factor $\alpha^2$ (with $\alpha$ the electromagnetic fine structure constant) compared to tree-level processes. Also current detectors have an energy resolution of not better than 10 \%, so the signal will be smeared and difficult to detect against background processes. 

\begin{figure*}[t!]
  \begin{minipage}[t]{0.999\textwidth}
\begin{center}
  \includegraphics[width=0.49\textwidth] {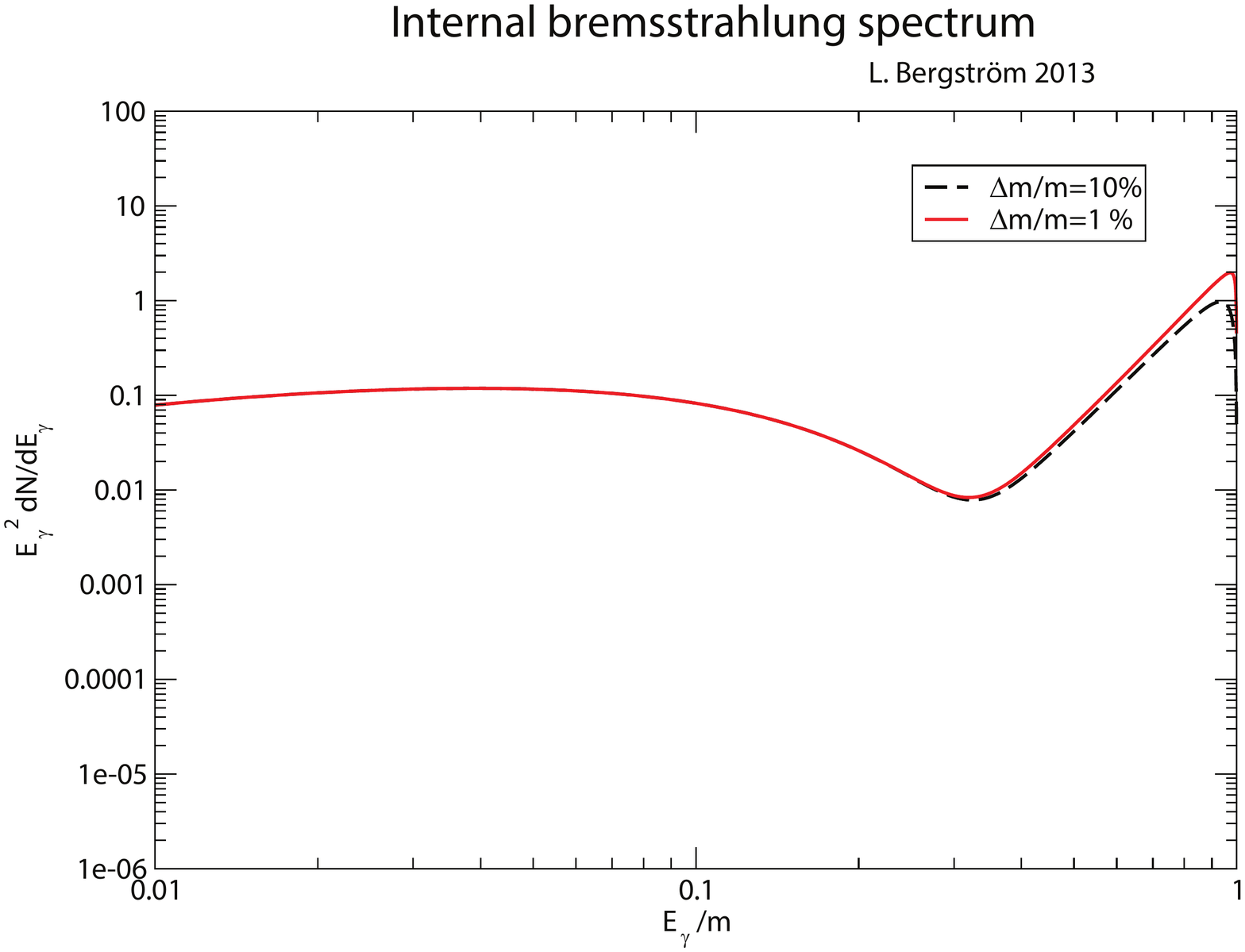}
\includegraphics[width=0.49\textwidth] {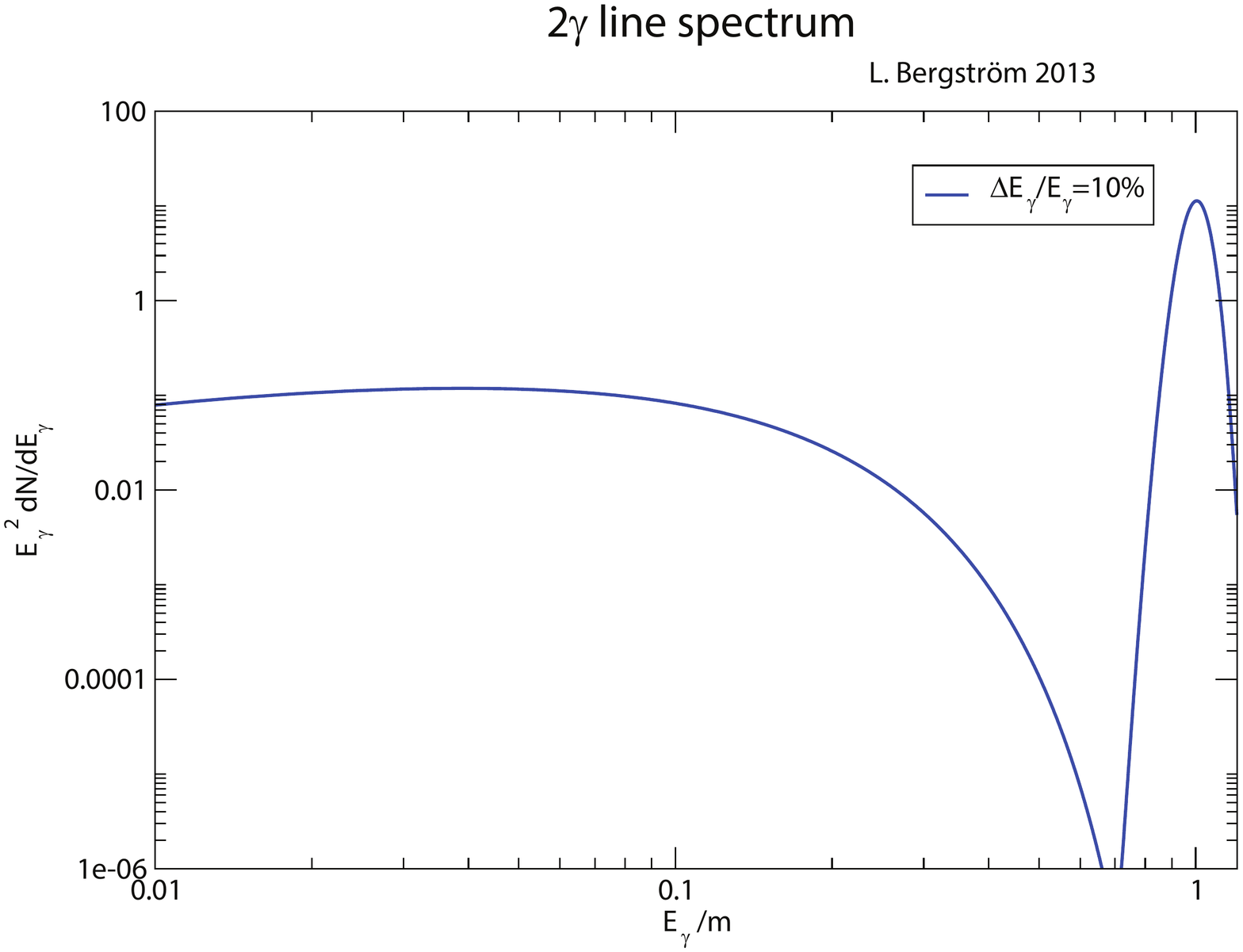}
\end{center}   
\end{minipage}
\caption{{\em Example of internal bremsstrahlung spectrum (left figure) and 
2$\gamma$ line spectrum (right figure). For both processes a typical continuum 
spectrum from quark fragmentation has been included. For the line an arbitrary relative normalization of the two has been assumed, and a 
10 \% energy resolution of the detector. As can be seen, both processes
 give a very characteristic rise in the $E^2dN/dE$ distribution near the
 kinematical endpoint $E/m=1$. Figure based on \protect\cite{annalen}.}}
\label{fig:spectra}
\end{figure*}

There is, however, another 
process, only suppressed linearly in $\alpha$ \cite{lbe89,bbe} and which is also peaked 
near the maximal photon energy  $E_\gamma=m_\chi$, and may be especially important for Majorana DM. This is the so-called internal bremsstrahlung process, which circumvents the helicity suppression  \cite{goldberg} for slow, annihilating Majorana particles.
The spectral shape of the emitted photon is favourable for detection as it increases rapidly with photon energy to give a peak close to the DM particle mass. 
Examples of these 
rather striking energy distributions are given in
Fig.~\ref{fig:spectra}.

Recently, these radiative processes have been generalized 
also to emission of other gauge bosons, and have been shown to be quite 
important generally \cite{radiative}. In particular, radiation of $W$ and $Z$ bosons can give observable antiproton rates for models that explain the PAMELA positron excess \cite{garny}.

The dependence on the square of the number density is an important feature of indirect detection. As numerical 
simulations have discovered that galactic halos may be quite crowded with 
sub-halos (see, e.g., \cite{springel}), this means that there is a possibility that the entirety of dark 
matter in the universe by means of this substructure in the universe 
may generate a cosmological signature 
\cite{cosmic}.  
 
Recently, indirect methods have started to be a competitive and complementary alternative to direct detection. The Fermi collaboration has focused on the nearest substructures, dwarf galaxies, that may be the visible counterpart of the ``clumpy'' dark matter seen in DM only simulations. The first limits on the interesting WIMP region were obtained by stacking data from several dwarf galaxies \cite{maja} (see also \cite{koushiappas}).

Presently, indirect $\gamma$-ray DM detection is in rapid evolution,
in particular thanks to the successful  Fermi-LAT space 
detector, and also the plans for a very large imaging air Cherenkov
telescope array, CTA \cite{CTA} which will follow the successful detectors HESS \cite{hess}, MAGIC \cite{magic} and VERITAS \cite{veritas}.

By analyzing public Fermi-LAT data, an excess in the few GeV $\gamma$-ray 
energy region has been proposed to be caused by DM annihilation. 
This $\gamma$-ray  diffuse emission comes from the  the region of the galactic centre 
(see, e.g., \protect\cite{hooper_12}).
This possible DM  effect (with a DM mass of order 10 GeV) is unexplained at the moment, but 
astrophysical explanations are of course possible in this crowded part of the galaxy.  It may be noted that the analysis of AMS-02 positron data \cite{lbe_hooper} would put some strain on some of the possible WIMP models proposed for the excess.

\subsection{The Mystery of the  $\gamma$ Line}
Recently, an interesting DM feature  was found \cite{bringmann,weniger,tempel,finkbeiner}, using public Fermi-LAT data \cite{fermi_lat}.
A structure in the energy spectrum consistent with  internal bremsstrahlung,
 or alternatively a narrow $\gamma$-ray line, was seen to be visible in the
 energy range around 150 GeV, for bremsstrahlung, or 130 GeV for the $\gamma\gamma$ line interpretation. 
(The spectrum may also be a combination of the two, and a weaker $Z\gamma$ line at lower energy may also be present \cite{lbe12}.)

The line feature has also been studied by the Fermi-LAT collaboration itself \cite{elliott}, who are however puzzled by the effect. First, they have an unexplained background at similar energy from $\gamma$-rays generated by cosmic rays hitting the atmosphere (and which thus can not have anything to do with dark matter). It also seems that a reprocessing of the data diminishes the significance of the peak. With Fermi-LAT taking data for several more years, it will 
be interesting to follow this development. In fact, the collaboration has recently decided to change the observation strategy from the beginning of 2014 to collect more data from the galactic centre region.

At energies, say, below 150 GeV, the Fermi-LAT instrument will in terms of energy resolution be superior to imaging air Cherenkov telescopes like the CTA, 
which will have its greatest sensitivity in the TeV range. However, new space detectors are currently being planned with a tenfold increase in energy resolution, an example  
being   GAMMA-400
 \cite{gamma-400}, which has a launch from Russia around 2018 approved. It is planned 
to have a slightly smaller effective area than Fermi-LAT, but with improved angular 
resolution. It will also have 
better energy resolution than Fermi-LAT by  an order of magnitude. 
This would take the search for unique $\gamma$-ray signatures of DM to another level of sensitivity. 
If the present indication of a line signal would persist, it should be seen 
in GAMMA-400 with a significance of the order of $10\sigma$ \cite{bbcfv}. It turns out, however, that in most models the rate is a factor of 3 - 10 smaller (see, e.g., \cite{lbe12}), but this could still be detectable, over a large energy range. There is also a similar project planned by Chinese scientists \cite{dampe}. 
 
Of course, the tentative line seen in Fermi-LAT data may turn 
out to be a statistical fluctuation or due to an (unknown) instrumental effect. To be convinced 
that the DM problem has found a solution, several independent
measurements will be needed, using all three methods we have at our disposal:
accelerator searches, direct and indirect detection, all having
a nice complementarity.  
\section{Concluding Remarks}

As search methods for dark matter now have matured, several experiments are currently surveying
regions of parameter space where a signal {\em may} be found. The first evidence will probably be of low significance -- meaning that false alarms have and will most certainly appear. However, with more time, and using all methods at our disposal, the probability of finally identifying dark matter, presently the only really serious problem for the Standard Model of particle physics to explain, will steadily increase. There is a bright future for dark matter searches!    
\section*{Acknowledgements}
I wish to thank the organizers, led in an eminent way by Tord Ekel\"of, for 
a very enjoyable and useful Nobel Symposium.
This work is supported by the Swedish Research Council under grant no. 621-2012-2250.

\end{document}